\documentclass[fleqn,usenatbib]{mnras}

\usepackage{newtxtext,newtxmath}
\usepackage[T1]{fontenc}

\DeclareRobustCommand{\VAN}[3]{#2}
\let\VANthebibliography\thebibliography
\def\thebibliography{\DeclareRobustCommand{\VAN}[3]{##3}\VANthebibliography}

\usepackage{graphicx}	% Including figure files
\usepackage{amsmath}	% Advanced maths commands
\usepackage{glossaries}

\newacronym{ao}{AO}{adaptive optics}
\newacronym{mcao}{MCAO}{multi conjugate adaptive optics}
\newacronym{ltao}{LTAO}{laser tomographic adaptive optics}
\newacronym{eso}{ESO}{European Southern Observatory}
\newacronym{shimm}{24hSHIMM}{24-hour Shack-Hartmann image motion monitor}
\newacronym{fass}{FASS}{full aperture scintillation sensor}
\newacronym{ringss}{RINGSS}{ring-image next generation scintillation sensor}
\newacronym{scidar}{S-SCIDAR}{Stereo-Scintillation Detection and Ranging}
\newacronym{scidar1}{SCIDAR}{Scintillation Detection and Ranging}
\newacronym{mass}{MASS}{Multi Aperture Scintillation Sensor}
\newacronym{mass-dimm}{MASS-DIMM}{Multi Aperture Scintillation Sensor - Differential Image Motion Monitor}
\newacronym{ot}{OT}{optical turbulence}
\newacronym{dimm}{DIMM}{Differential Image Motion Monitor}
\newacronym{asm}{ASM}{Atmospheric Site Monitoring}
\newacronym{aa}{AoA}{angle-of-arrival}
\newacronym{shwfs}{SHWFS}{Shack-Hartmann wavefront sensor}
\newacronym{sct}{SCT}{Schmidt-Cassegrain Telescope}
\newacronym{slodar}{SLODAR}{Slope Detection and Ranging}
\newacronym{elt}{ELT}{Extremely Large Telescope}
\newacronym{scao}{SCAO}{single conjugate adaptive optics}
\newacronym{vlt}{VLT}{Very Large Telescope}
\newcommand{\cn}{\ensuremath{C_n^2(h)\:\mathrm{d}h} } % 

\title[Comparing next-generation turbulence profilers at Paranal]{A comparison of next-generation turbulence profiling instruments at Paranal}

\author[]{
\newauthor R. Griffiths,$^{1}$ \thanks{E-mail: ryan.griffiths@durham.ac.uk}
L. Bardou,$^{1}$
T. Butterley,$^{1}$
J. Osborn,$^{1}$
R. Wilson,$^{1}$
E. Bustos,$^{2}$
A. Tokovinin,$^{2}$
\newauthor M. Le Louarn,$^{3}$
and A. Otarola $^{4}$ \\
$^{1}$Centre for Advanced Instrumentation, University of Durham, DH1 3LE, Durham, UK\\
$^{2}$Cerro Tololo Inter-American Observatory — NSF’s NOIRLab, Casilla 603, La Serena, Chile\\
$^{3}$European Southern Observatory, Karl-Schwarzschild-Str. 2 85748 Garching bei München Germany\\
$^{4}$European Southern Observatory, Alonso de Córdova 3107 Vitacura, Santiago de Chile, Chile}

\date{Accepted XXX. Received YYY; in original form ZZZ}

\pubyear{2023}

\begin{document}
\label{firstpage}
\pagerange{\pageref{firstpage}--\pageref{lastpage}}
\maketitle

\begin{abstract}
{A six-night optical turbulence monitoring campaign  has been carried at Cerro Paranal observatory in February and March, 2023 to facilitate the development and characterisation of two novel atmospheric site monitoring instruments - the ring-image next generation scintillation sensor (RINGSS) and 24-hour Shack Hartmann image motion monitor (24hSHIMM) in the context of providing optical turbulence monitoring support for upcoming 20-40m telescopes. Alongside these two instruments, the well-characterised Stereo-SCIDAR and 2016-MASS-DIMM were operated throughout the campaign to provide data for comparison. All instruments obtain estimates of optical turbulence profiles through statistical analysis of intensity and wavefront angle-of-arrival fluctuations from observations of stars. Contemporaneous measurements of the integrated turbulence parameters are compared and the ratios, bias, unbiased root mean square error and correlation of results from each instrument assessed. Strong agreement was observed in measurements of seeing, free atmosphere seeing and coherence time. Less correlation is seen for isoplanatic angle, although the median values agree well. Median turbulence parameters are further compared against long-term monitoring data from Paranal instruments. Profiles from the three small-telescope instruments are compared with the 100-layer profile from the stereo-SCIDAR. It is found that the RINGSS and SHIMM offer improved accuracy in characterisation of the vertical optical turbulence profile over the MASS-DIMM. Finally, the first results of continuous optical turbulence monitoring at Paranal are presented which show a strong diurnal variation and predictable trend in the seeing. A value of 2.65$\arcsec$ is found for the median daytime seeing.}

\end{abstract}


%%%%%%%%%%%%%%%%%%%%%%%%%%%%%%%%%%%%%%%%%%%%%%%%%%

%%%%%%%%%%%%%%%%% BODY OF PAPER %%%%%%%%%%%%%%%%%%

\section{Introduction}

Atmospheric \gls{ot} induces both phase distortion and amplitude modulation of light that propagates through it, leading to a severe reduction in achievable image quality from ground-based optical instruments. Large astronomical telescopes typically employ \gls{ao} systems to compensate for the wavefront phase distortion, however there is a need for external monitoring of \gls{ot} during the design, validation and commissioning of such systems. Additionally, knowledge of the vertical distribution of optical turbulence will be crucial for predicting and verifying the performance of \gls{mcao} systems planned for 20-40m ELT-class telescopes \citep{Costille2011, Tokovinin2010}. These systems will therefore demand instruments that measure both "integrated" parameters relevant to \gls{ao} and the vertical distribution of optical turbulence. Turbulence monitoring instruments are today installed at many of the largest astronomical observatories, providing real-time measurements of turbulence conditions, ensuring that observational sensitivity requirements are met \citep{Milli2019NowcastingObservatory}, and providing long-term site monitoring data which is highly desirable in the development of new optical instruments. Turbulence monitoring is also seen as increasingly important in improving the accuracy of meso-scale turbulence forecasting models \citep{Masciadri2019}, which offer further gains in efficiency for observation scheduling through the process of auto-regression \citep{Masciadri2023OpticalTelescope} and will be highly beneficial to the operation of ELT-class instruments. The current standard, small-telescope \gls{ot} monitoring instruments - the \gls{mass} and \gls{dimm} - are limited by the use of outdated CCD cameras, custom-manufactured equipment and, in the case of the \gls{mass}, a noted discrepancy in measurements of \gls{ot} profiles compared to the high-resolution \gls{scidar} technique \citep{Masciadri2014OnTechniques, Lombardi2016UsingTechniques}. There is therefore significant motivation to develop new instruments based on modern technologies for deployment alongside ELTs.

The minimum requirement for such instruments is firstly accurate measurement of the astronomical seeing $\varepsilon_0$. This parameter is directly related to the integrated turbulence strength of the atmosphere and represents the angular size of the seeing-limited (long-exposure) point spread function (PSF) for astronomical observations. The free atmosphere seeing, $\varepsilon_{0,f}$ is a measure of the seeing above an altitude of 500m \citep{Lawrence2004ExceptionalAntarctica} and enables a comparison of seeing decoupled from highly localised turbulence in the ground layer. Additional integrated turbulence parameters of interest include the coherence time, $\tau_0$, and isoplanatic angle, $\theta_0$ \citep{Roddier1981}. These are relevant to the operation of \gls{ao} systems, representing respectively an upper limit on the time taken to measure and correct wavefront distortions and an upper limit of the achievable angular correction. \gls{mcao} and \gls{ltao} systems planned for ELT-instruments will also require knowledge of the optical turbulence profile, as do forecasting models, in order to provide meaningful validation of techniques. Accurate measurement of the optical turbulence profile is therefore also highly desirable. 

Multi-instrument campaigns have been hosted a number of times at the \gls{eso} Paranal site, including for example \citet{DaliAli2010} and \citet{Osborn2018Profiling}. This work details the results from the most recent campaign at Paranal, in which three turbulence profiling instruments based on portable telescopes: the \gls{shimm} \citep{Griffiths2023}, \gls{fass} \citep{Guesalaga2021FASS:Camera} and \gls{ringss} \citep{Tokovinin2021MeasurementImages} were compared with permanently installed \gls{ot} profiling instruments at the site. The primary motivation being to facilitate the development and characterisation of these next-generation instruments against existing techniques. The three instruments were co-located on the northernmost part of the observatory for 6 nights starting on the 27th of February, with the final night of observation on the 5th of March 2023. The ESO \gls{mass-dimm} \citep{Chiozzi2016TheUpgrade} was operating throughout all nights of observation whereas the stereo-SCIDAR \citep{Osborn2018} was operated from the 28th to the 5th only. As a part of the VLT \gls{asm} package, measurements of local meteorological parameters were available for additional analysis.

This work will outline the theoretical operating principle behind each instrument used in the campaign and present the major results from the campaign with discussion. The generalised \gls{fass} instrument is still under development and so its results have been excluded from this work. The measurements of the \gls{shimm} and \gls{ringss} will be compared directly with the permanent instrumentation - the \gls{dimm}, \gls{mass-dimm} and the \gls{scidar} - both on measurements of integrated parameters and on \gls{ot} profiles using high-resolution vertical $C_n^2$ profiles obtained from the \gls{scidar}.

\section{Turbulence profiling instruments}\label{section:2}

The concepts and capabilities of each of the instruments used during the campaign are briefly summarised below. For this campaign, the other ESO turbulence profiling instruments: the robotic \gls{slodar} instrument and the adaptive optics facility on UT-4, were not operational and so are omitted.

\subsection{Stereo-SCIDAR}

\gls{scidar}, which is described in detail in \cite{Shepherd2014}, is a triangulation technique that exploits observations of binary stars with a similar magnitude, requiring a telescope larger than 1-m diameter and low-noise camera due to the relative faintness of such targets, to measure the vertical distribution of \gls{ot} in the atmosphere. The \gls{scidar} projects the pupil image from each star onto a separate CCD detector using a prism which yields sensitivity advantages over the typical SCIDAR implementation where the pupil images are overlapped on a single camera \citep{Fuchs1998FocusingSCIDAR}. The cross covariance of the spatial intensity fluctuations in the two pupil images is analysed to extract a high-resolution optical turbulence \cn profile comprised of 100 layers at 250m intervals. Additionally, by analysing the temporal evolution of the cross-covariance responses, it is possible to extract the wind velocity and direction of individual turbulent layers which enables estimation of the optical turbulence coherence time. The \gls{scidar} system at Paranal is mounted on one of the 1.8m auxiliary telescopes and has been extensively tested and validated against existing instrumentation at the site \citep{Osborn2018Profiling}. The \gls{scidar} data from this experiment has been processed using the latest corrections for finite spatial sampling described by \citet{Butterley2020CorrectionStereo-SCIDAR} which also includes subtraction of localised turbulence within the dome.

\subsection{DIMM}
The \gls{dimm} \citep{SARAZIN1990} consists of a small telescope with a CCD camera and a pupil-plane mask of two small circular apertures. Using a prism, the beams from the two apertures are imaged onto a detector and spatially separated. The seeing is measured by analysing the variance in differential position of the two focal spots \citep{Tokovinin2002}. The \gls{dimm} is a simple, portable \gls{ot} monitor and provides measurements of the seeing at one minute intervals. The \gls{vlt} \gls{dimm} at Paranal is configured in a combined \gls{mass-dimm} system mounted on a 28-cm Celestron C11 telescope and was installed as a part of the 2016 \gls{asm} upgrade on a 7-m tower. Limitations of the instrument include insensitivity to the bias introduced by optical propagation and only providing measurements of the seeing.

\subsection{MASS}\
The \gls{mass} \citep{Kornilov2003MASS:Distribution} is similarly based around a small-telescope and measures the normalised intensity fluctuations resulting from propagation though turbulence, commonly referred to as the scintillation index, in 4 concentric apertures. Using the theory described by \citet{Tokovinin2003RestorationIndices}, weighting functions are generated for the 10 (4 normal and 6 differential) scintillation indices at vertical heights of 0, 0.5, 1, 2, 4, 8, 16 km and an inversion algorithm is used to reconstruct the $C_n^2(h)\,\mathrm{d}h$ of each layer. The \gls{vlt} \gls{mass} is combined in a \gls{mass-dimm} configuration \citep{Kornilov2007CombinedStudies}. As the \gls{mass} relies solely on measurements of scintillation, it is insensitive to ground-layer turbulence which can be accounted for using simultaneous measurements from the \gls{dimm}. The techniques described by \cite{Kornilov2011StellarEvaluation} allow for estimation of the \gls{ot} coherence time by measurement of the atmospheric second moment of wind and combination with the \gls{dimm} data.

\subsection{RINGSS}\label{sec:ringss}

RINGSS is a solid-state turbulence profiler developed to replace the technically obsolete MASS instruments \citep{Tokovinin2021MeasurementImages}. It uses a 5-inch Celestron telescope where image of a bright single star is optically transformed into a ring. This is achieved by combination of spherical aberration and defocus in the focal-reducer lens. The pixel scale is 1.57 arcsec and the ring radius is 11 pixels. Cubes of 2000 ring images of 48$\times$48 pixel format and 1 ms exposure time are recorded by a CMOS camera. Image processing consists in centering the rings and computing 20 harmonics of intensity variation along the ring (in the angular coordinate). Variances of these harmonics, averaged over 10 image cubes, are related to the turbulence profile by means of weighting functions in the same way as in MASS. RINGSS delivers turbulence integrals in eight  layers at 0, 0.25, 0.5, 1... 16 km heights. The results refer to zenith; they are corrected for the  finite exposure time bias and partially corrected for deviations from the weak-scintillation regime (saturation). The atmospheric time constant is determined by the method of \citet{Kornilov2011StellarEvaluation}. The instrument operates robotically. Its control provides for selection and change of targets, pointing and centering, and closed-loop focus control.

Scintillation signals in RINGSS are sensitive to the ground-layer turbulence because the image is not focused (analogue of a generalized SCIDAR). Alternative estimation of seeing is made using radial distortions of the rings, like in a DIMM. This  "sector" seeing 
agrees reasonably well with the scintillation-based seeing: the ratio of their mean values is 1.038, the correlation coefficient is 0.97, and the rms scatter around the regression line is 0.11$''$. Under excellent conditions, the sector seeing is systematically larger; this bias appears when turbulence in the ground layer is less than $2 \times 10^{-13}$ m$^{1/3}$ and is absent otherwise. We attribute this effect to imperfect focusing of the ring in the radial direction, analogous to the similar bias in a defocused DIMM. In the following analysis, we use only the scintillation-based seeing measured by RINGSS, while the supplementary data  provide the alternative "sector" seeing values as well.

\subsection{24hSHIMM}\label{sec:shimm}
The \gls{shimm} \citep{Griffiths2023} is based around a \gls{shwfs} and portable 11-inch telescope design. It observes single, bright stars and measures both the intensity and wavefront \gls{aa} fluctuations in each of the \gls{shwfs} focal spots. The spatial statistics of the scintillation are compared with weighting functions \citep{Robert2006} and a non-negative least squares algorithm is used to reconstruct a low-resolution $C_n^2(h)\, \mathrm{d}h$ profile. The \gls{shimm} is not negatively-conjugated, therefore a scintillation-based reconstruction is insensitive to the ground layer and integrated turbulence strength measurements from \gls{shwfs} \gls{aa} fluctuations are used to overcome this limitation. The \gls{shimm} is designed to operate for 24-hours a day, typically through use of an InGaAs camera operating in the short-wave infrared to reduce sky background light and minimise the effects of strong turbulence. The \gls{shimm} utilises the FADE method \citep{Tokovinin2008FADETime} to estimate the coherence time of the atmospheric turbulence. This method of direct measurement of coherence time is an improvement on the previous implementation using wind-speed profiles from the ERA5 ECMWF forecast \citep{Hersbach2020} which are limited by low spatial and temporal resolution.
%The \gls{shimm} does not operate at high frame rates to measure the coherence time, however instead it uses the estimate of the turbulence profile along with wind speed measurements from a local weather station and the vertical wind speed profile from the ERA5 ECMWF forecast \citep{Hersbach2020} to obtain an estimate of the coherence time. 
Another notable change from the original implementation of the \gls{shimm} is that in this work, measurements are obtained by a CMOS camera and a 600nm longpass filter which introduces additional constraints on performance.

\subsection{Campaign details}\label{sec:method}
\begin{figure}
\centering
\includegraphics[width=\hsize, trim = 0cm 2cm 14cm 1cm]{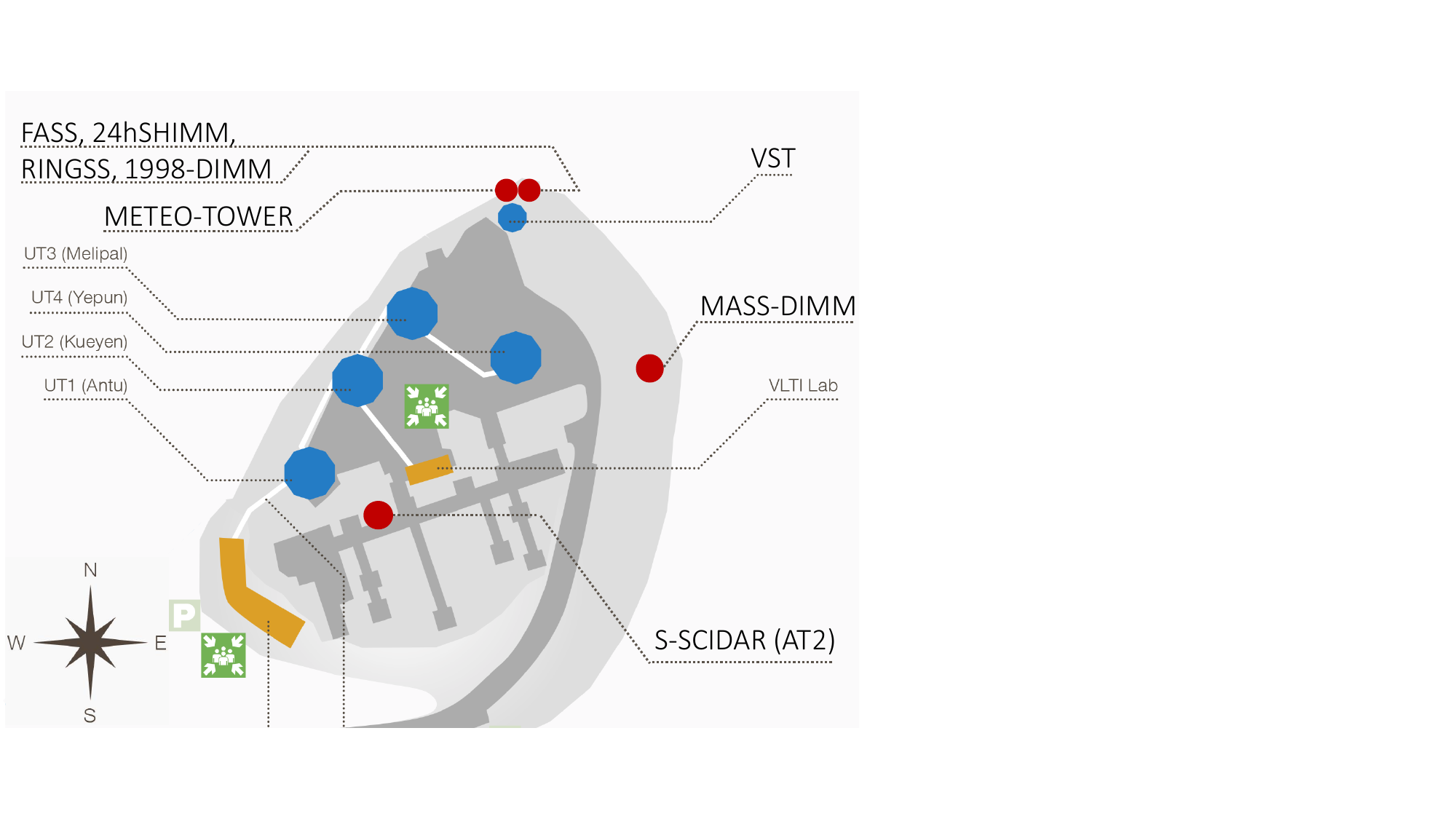}
\caption{Location of turbulence monitoring instrumentation described in section \ref{section:2}. Instruments relevant to this study are indicated by red circles. Original image credit: ESO.} 
\label{fig:platform}
\end{figure}

The location of each instrument on the Paranal observatory platform is shown in figure ~\ref{fig:platform}. The \gls{shimm} and \gls{ringss} were mounted on concrete pillars adjacent to the 1998 \gls{dimm} tower within 2m of one-another. The \gls{fass} was mounted on a tripod slightly further away, between the old-\gls{dimm} tower and \gls{slodar} crate. The \gls{shimm} was mounted approximately 2m off of the ground, the \gls{ringss} and \gls{fass} were at about 1.5m. Wind breaks were set up along the Northern fence next to the instruments.

The local environments for the \gls{scidar} and \gls{mass-dimm} are therefore significantly different; they are both much further away from any large buildings and more elevated from the ground. The \gls{mass-dimm} is on a 7~m tower and the \gls{scidar} was mounted on VLTI auxiliary telescope two; the alt-az altitude axis of which is 5-m above surface \citep{Koehler2000VLTITelescopes}. We therefore expect poorer agreement in the seeing between these instruments and the monitors located near the VLT Survey Telescope (VST), as local turbulence conditions are likely to differ significantly. 

The list of targets for the \gls{ringss} was shared at the beginning of the experiment and efforts were made to synchronise target stars where possible between the visiting turbulence monitors. The \gls{mass-dimm} and \gls{scidar} however were using separate target lists. 

\section{Results}

The overall results for this campaign are laid out below. This includes both direct comparison of integrated parameter measurements between the different instruments, and a comparison of optical turbulence profiles with the high-resolution \gls{scidar}. A focus is primarily made on comparison of the developmental instruments \gls{shimm}, \gls{ringss} with the well-characterised and permanently installed \gls{scidar} and \gls{mass-dimm}. However all instruments have been compared where appropriate. The comparison between \gls{shimm} and \gls{ringss} is of interest as the two instruments were co-located, observing similar targets and so are much more likely to agree. The agreement of the \gls{scidar} and \gls{mass-dimm} is also of interest to compare to long-term monitoring results and previous studies.

To generate comparison plots, for the instrument on the x-axis, each measurement has been directly plotted against the nearest measurement from the instrument on the y-axis within a maximum time difference of two minutes. If a corresponding measurement could not be found within two minutes, the data point has been excluded from the plot to minimise the effects of temporal evolution of the turbulence on the comparison. This two minute interval was chosen to match the integration time used by the \gls{scidar} as it was the longest of all the instruments. As the algorithm finds the nearest measurement within the search window and the other instruments all have a cadence of a minute or less, reducing the interval to one minute, for example, was observed to produce almost identical statistical comparison parameters.
%Additionally, each of the instruments has a similar integration time of approximately 20-120 s which allows for direct comparison between data packets. 
In each comparison plot, a white dashed line represents the line of perfect agreement between the instruments, and the Pearson correlation coefficient, $r$, bias, $\mathrm{B}$, unbiased root mean square error, $\mathrm{RMSE}$, and mean ratio, $\mathrm{MR}$, of each data set is reported in the top-left of the graph. Mathematical definitions of the latter three parameters may be found in appendix~\ref{sec:appendA}. These comparison parameters are additionally summarised for each figure in table \ref{tab:statistical_comparisons}. The colour gradient indicates the density of measurements at each point in the graph with black the lowest and pale yellow the highest. The median values from these findings will also be compared where useful to results from long-term studies on seeing conditions at Paranal with \citet{Butterley2021SCIDAR2b} reporting the latest \gls{scidar} results and \citet{Otarola2021} the results from the \gls{mass} and \gls{dimm}. These results can be found in table~\ref{table:All}. 
All integrated turbulence parameter measurements displayed below are derived at zenith and a wavelength of 500~nm. All turbulence profiles are given as a function of vertical height. Finally, the distribution and temporal sequences of \cn profiles measured by the instruments will be directly compared with the \gls{scidar} through a binning process to investigate accuracy of \gls{ot} profile characterisation, and the first results from the \gls{shimm} of 24-hour continuous monitoring of \gls{ot} at Paranal are presented in full.

\begin{table*}
\caption{Median values of parameters obtained during this campaign, marked in the columns as `2023', from all instruments are compared with long-term site monitoring results of \citet{Otarola2021, Butterley2021SCIDAR2b} with the column labels `long-term'. There are some blank entries which correspond to unavailable data - either because the instrument cannot measure the parameter or there is no source for long-term data. The median values for the \gls{shimm} are calculated excluding data taken during the daytime.}             
\label{table:All}      
\centering          
\begin{tabular}{l l c c c c c c c c c}     % 7 columns 
\hline\hline       
                      % To combine 4 columns into a single one 
& N Profiles & \multicolumn{2}{c}{$\varepsilon_0 (\arcsec)$} & \multicolumn{2}{c}{$\varepsilon_{0,f} (\arcsec)$} & \multicolumn{2}{c}{$\theta_0 (\arcsec)$} & \multicolumn{2}{c}{$\tau_0$ (ms)} \\
\hline
Instrument & 2023 & Long-term & 2023 & Long-term & 2023 & Long-term & 2023 & Long-term & 2023 \\ 
\hline     
   \gls{dimm} & 2696 & 0.71 & 0.75 & - & - & - & - & - & - &  \\
   \gls{mass-dimm} & 2477 & - & 0.79 & 0.41 & 0.40 & 1.98 & 2.53 & 6.14 & 6.3\\
   \gls{scidar} & 611 & 0.72 & 0.76 & 0.46 & 0.51 & 2.03 & 2.62 & 3.61 & 5.8\\  
   \gls{ringss} & 5387 & - & 1.10 & - & 0.58 &  & 2.46 & - & 5.8\\
   \gls{shimm} & 1942 & - & 0.89 & - & - & - & 2.35 & - &  6.4\\
\hline                  
\end{tabular}
%\tablefoot{\tablefoottext{a}{Night-time profiles only, total profiles 3712.}}
\end{table*}

\subsection{Seeing}\label{sec:seeing}

The astronomical seeing, $\varepsilon_0$, describes the angular full-width-at-half-maximum (FWHM), typically measured in units of arcseconds, of the seeing-limited point spread function for long-exposure imaging through optical turbulence. It can be calculated using the Fried parameter $r_0$ \citep{Fried1966OpticalExposures},
\begin{equation}
    r_0 = \left[0.423 k^2 \sec(\gamma) \int^{\infty}_0 C_n^2(h)\, \mathrm{d}h \right] ^{-3/5},
\end{equation}
where $k = 2\pi/\lambda$ is the wavenumber, $\lambda$ is the wavelength of the light, $\gamma$ the zenith angle of observation in radians, $h$ the altitude of a turbulent layer in metres, $C_n^2(h)$ the refractive index structure constant, given in units of $\mathrm{m}^{-2/3}$. The relationship between the Fried parameter and the seeing is then given by
\begin{equation}
    \varepsilon_0 = 0.98 \frac{\lambda}{r_0}.
\end{equation}

Accurate measurement of the astronomical seeing is the most fundamental requirement of an optical turbulence monitor as it quantifies the integrated turbulence strength of the atmosphere and directly relates this to the degree of image distortion. 
Seeing is dynamic, can change rapidly and is highly dependant on location and pointing direction \citep{Tokovinin2023TheSeeing} which leads to discrepancies between instruments, even for well-synchronised measurements.
%\textcolor{red}{Median seeing measurements of 1.10\arcsec and 0.89\arcsec from \gls{shimm} and \gls{ringss} compared with 0.75\arcsec and 0.76\arcsec found by the \gls{dimm} and \gls{scidar}}
Median seeing measurements in table~\ref{table:All} indicate that the two instruments located in the northern end of the site, near to the VST and installed at a lower height above ground, are measuring substantially stronger seeing that the \gls{mass-dimm} and \gls{scidar}. This is most likely due to local turbulence effects. There is however a very strong agreement between the \gls{dimm} and \gls{scidar} measurements, and a mean ratio close to 1, despite their separation on the site --- but noting their similar height above the ground and isolated locations this is not surprising.  

It is known that the local seeing at the 1998-DIMM tower is slightly stronger than the current 2016-\gls{mass-dimm}. The median seeing calculated from several years of measurements with the 1998-DIMM between 2010-01-01 and 2015-05-22 was found to be 0.98$\arcsec$ compared to the 2016-DIMM long term seeing of 0.71$\arcsec$. %\textcolor{red}{More recently it has been demonstrated that poor correlation can be found between instruments at different locations at the site, with a correlation of 0.62 found between the current \gls{dimm} and the NAOMI adaptive optics system on the ATs \citep{Morujao2023IntegratedData}}. 
This supports a location-based argument for some of the discrepancy between the visiting and the \gls{eso} instruments. Previous campaigns using the Generalised Seeing Monitor at the same location have found seeing values of 0.88$\arcsec$ \citep{Martin2000OpticalContribution} and 1.07$\arcsec$ \citep{DaliAli2010}. Additionally, high-resolution profiling of the surface layer carried out by \cite{Butterley2020CharacterizationSystem} using the surface-layer \gls{slodar} identifies an exponentially decaying turbulence strength with altitude --- hence we also expect the higher elevation of the \gls{mass-dimm} and \gls{scidar} to result in lower seeing.

\begin{figure}
\centering
\includegraphics[width=\hsize]{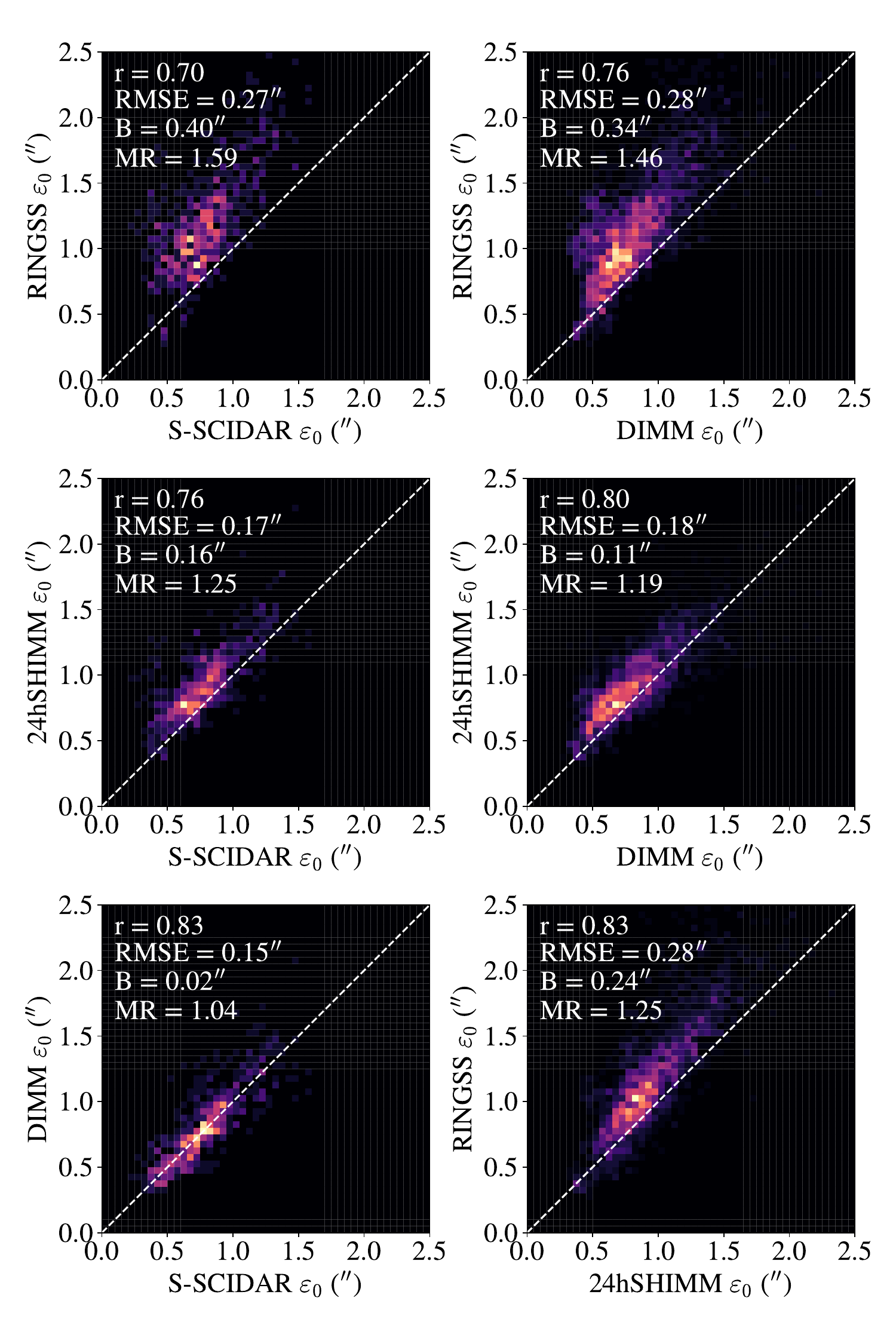}
\caption{Comparison of contemporaneous seeing measurements during the campaign from the \gls{dimm}, \gls{scidar}, \gls{shimm} and \gls{ringss}.}
\label{fig:seeing}
\end{figure}

Individual comparisons of seeing measured by each instrument are displayed in figure \ref{fig:seeing}. It is extremely encouraging that all seeing measurements display strong correlation with the minimum of $r = 0.70$ for the \gls{ringss} compared with the \gls{scidar}. As expected, due to co-location and overlapping targets, the \gls{shimm} and \gls{ringss} display a very strong correlation of 0.83, however there is a significant bias between the two despite their proximity. A number of factors may contribute to this, including the \gls{ringss} corrections for finite exposure time and partial saturation of scintillation - conditions which would lead to underestimates of fast-moving and high altitude turbulent layers on the \gls{shimm} - there is a also a small height offset between the two with the \gls{ringss} being closer to the ground which could lead to slightly stronger turbulence above the telescope pupil. The correlation between the \gls{dimm} and \gls{scidar} is equally strong but with far less bias - the results are also consistent with the long term monitoring 
%\textcolor{red}{with the median seeing measurements of 0.75\arcsec and 0.76\arcsec close to the corresponding long-term averages of 0.71\arcsec and 0.72\arcsec.} 
 as seen in table~\ref{table:All}.

%\begin{table}
%\caption{Summary of statistical comparison parameters for contemporaneous measurements of seeing.}
% \label{tab:seeing}
% \begin{tabular}{llcccc}
%  \hline
%   X - axis & Y - axis & r & RMSE & B & MR\\
%    & & & ($\arcsec$) & ($\arcsec$) &  \\
%    \hline
%  \gls{scidar} & \gls{ringss} & 0.70 & 0.49 & %0.40  & 1.59 \\
 % \gls{dimm} & \gls{ringss} & 0.76 & 0.44 & %0.34 & 1.46 \\
 % \gls{scidar} & \gls{shimm} & 0.76 & 0.23 & 0.16 & 1.25\\
 % \gls{dimm} & \gls{shimm} & 0.80 & 0.22 & 0.12 & 1.19\\
 % \gls{scidar} & \gls{dimm} & 0.83 & 0.16 & 0.02 & 1.04\\
 % \gls{shimm} & \gls{ringss} & 0.83 & 0.37 & 0.24 & 1.25\\
 % \hline
 %\end{tabular}
%\end{table}

\begin{table}
 \caption{Summary of statistical comparison parameters all graphs.}
 \label{tab:statistical_comparisons}
 \begin{tabular}{llcccc}
  \hline
   X - axis & Y - axis & r & RMSE & B & MR\\
    \hline
    \multicolumn{2}{c}{Seeing, $\varepsilon_0$} & & ($\arcsec$) & ($\arcsec$) &  \\
    \hline
  \gls{scidar} & \gls{ringss} & 0.70 & 0.27 & 0.40  & 1.59 \\
  \gls{dimm} & \gls{ringss} & 0.76 & 0.28 & 0.34 & 1.46 \\
  \gls{scidar} & \gls{shimm} & 0.76 & 0.17 & 0.16 & 1.25\\
  \gls{dimm} & \gls{shimm} & 0.80 & 0.18 & 0.11 & 1.19\\
  \gls{scidar} & \gls{dimm} & 0.83 & 0.15 & 0.02 & 1.04\\
  \gls{shimm} & \gls{ringss} & 0.83 & 0.28 & 0.24 & 1.25\\
  \hline
   \multicolumn{2}{c}{Free atmosphere seeing, $\varepsilon_{0,f}$} & & ($\arcsec$) & ($\arcsec$) &  \\
   \hline
  \gls{scidar} & \gls{ringss} & 0.86 & 0.21 & 0.14  & 1.23 \\
  \gls{mass-dimm} & \gls{ringss} & 0.85 & 0.22 & 0.14 & 1.36 \\
  \gls{scidar} & \gls{mass-dimm} & 0.80 & 0.17 & -0.03 & 0.93 \\
  \hline
     \multicolumn{2}{c}{Isoplanatic angle, $\theta_0$} & & ($\arcsec$) & ($\arcsec$) &  \\
    \hline
  \gls{scidar} & \gls{ringss} & 0.35 & 0.67 & -0.17  & 0.97 \\
  \gls{mass-dimm} & \gls{ringss} & 0.40 & 0.59 & -0.08 & 1.00 \\
  \gls{scidar} & \gls{shimm} & 0.40 & 0.67 & -0.32 & 0.91 \\
  \gls{mass-dimm} & \gls{shimm} & 0.33 & 0.65 & -0.18 & 0.96\\
  \gls{scidar} & \gls{mass-dimm} & 0.30 & 0.72 & -0.19 & 0.97\\
  \gls{shimm} & \gls{ringss} & 0.54 & 0.53 & 0.11 & 1.08\\
  \hline
   \multicolumn{2}{c}{Coherence time, $\tau_0$} & & (ms) & (ms) &  \\
    \hline
  \gls{scidar} & \gls{ringss} & 0.75 & 1.94 & -0.05  & 1.00 \\
  \gls{mass-dimm} & \gls{ringss} & 0.69 & 3.69 & -0.82 & 0.96 \\
  %\gls{scidar} & \gls{shimm} & 0.69 & 1.80 & -2.81 & 0.59 \\
  % \gls{dimm} & \gls{shimm} & 0.77 & 2.57 & -3.24 & 0.58\\
  \gls{scidar} & \gls{shimm} & 0.68 & 2.15 & 0.21 & 1.05 \\
   \gls{mass-dimm} & \gls{shimm} & 0.77 & 2.46 & 0.14 & 1.05\\
  \gls{scidar} & \gls{mass-dimm} & 0.70 & 2.10 & 0.25 & 1.05 \\
  \gls{shimm} & \gls{ringss} & 0.80 & 2.21 & -0.53 & 0.97 \\
  %\gls{shimm} & \gls{ringss} & 0.84 & 2.13 & 2.69 & 1.72 \\
  \hline
 \end{tabular}
\end{table}

\subsection{Free atmosphere seeing}\label{sec:freeatmos}

\begin{figure}
\centering
\includegraphics[width=0.5\hsize]{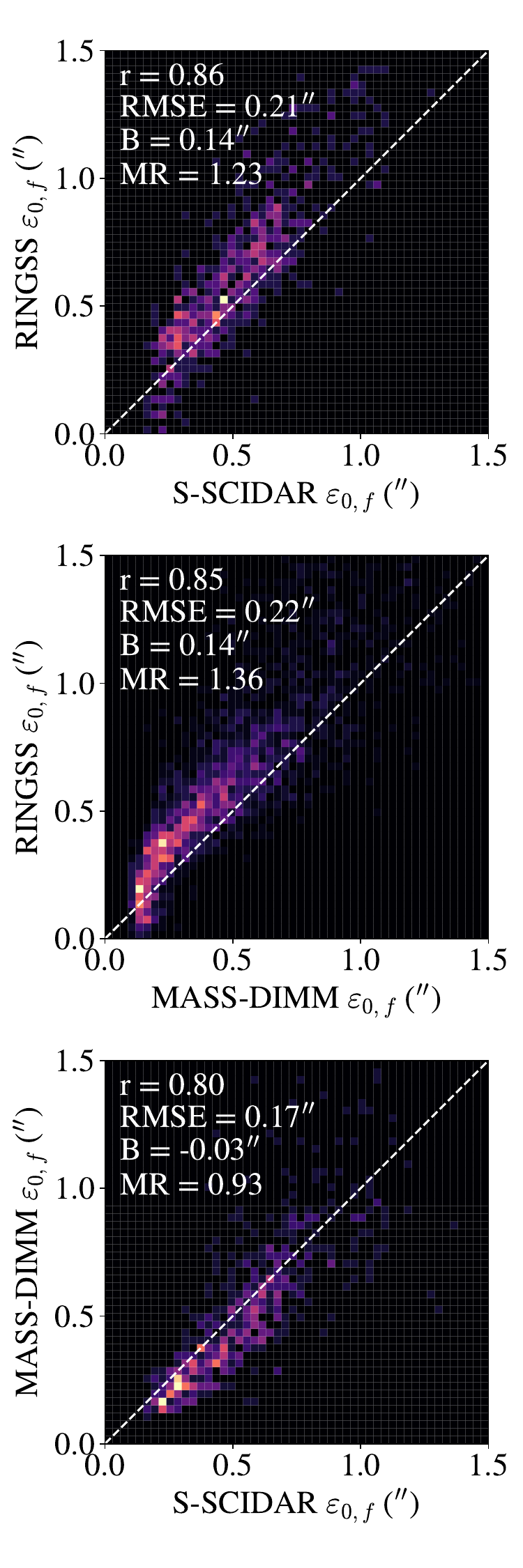}
\caption{Comparison of contemporaneous free atmosphere seeing measurements during the campaign from the \gls{mass-dimm}, \gls{scidar}, and \gls{ringss}.}
\label{fig:fsee}
\end{figure}

The free atmosphere seeing, $\varepsilon_{0,f}$ is calculated as the integrated seeing of all turbulent layers with an altitude of 500~m or greater for the \gls{mass}, \gls{ringss} and \gls{scidar}. The \gls{shimm} is limited by a large sub-aperture size of 4.7 cm and cannot sample the highest frequency scintillation fluctuations produced by low-altitude turbulence. This is due to height scaling of the characteristic size of scintillation speckles - given by the radius of the first Fresnel zone, $r\approx\sqrt{\lambda z}$. It therefore lacks the sensitivity required to reconstruct a layer at 500~m, so a direct comparison with the other instruments is not possible and it has been excluded. Figure~\ref{fig:fsee} details the measurements obtained with the three other instruments.

%\begin{table}
% \caption{Summary of statistical comparison parameters for contemporaneous measurements of free atmosphere seeing.}
% \label{tab:free_atmopshere_seeing}
% \begin{tabular}{llcccc}
%  \hline
%   X - axis & Y - axis & r & RMSE & B & MR\\
%   & & & ($\arcsec$) & ($\arcsec$) &  \\
%   \hline
%  \gls{scidar} & \gls{ringss} & 0.86 & 0.25 & 0.14  & 1.23 \\
%  \gls{mass-dimm} & \gls{ringss} & 0.85 & 0.26 & 0.14 & 1.36 \\
%  \gls{scidar} & \gls{mass-dimm} & 0.80 & 0.18 & -0.03 & 0.93 \\
%  \hline
% \end{tabular}
%\end{table}

\subsection{Isoplanatic angle}\label{sec:angle}

The isoplanatic angle is defined by \cite{Roddier1981} as
\begin{equation}\label{equation:isop}
    \theta_0 = \left[ 2.91 k^2 \cos^{-8/3}({\gamma}) \int_0^{\infty} C_n^2(h) \,h^{5/3} \mathrm{d}h \right]^{-3/5}.
\end{equation}

This quantity is of particular interest for design and operation of \gls{ao} systems as it represents the separation angle between a guide star and target which will result in 1~rad$^2$ RMS wavefront error for phase corrections. It is particularly of interest when considering target availability in \gls{scao} and in calculation of \gls{ao} error budgets.

\begin{figure}
\centering
\includegraphics[width=\hsize]{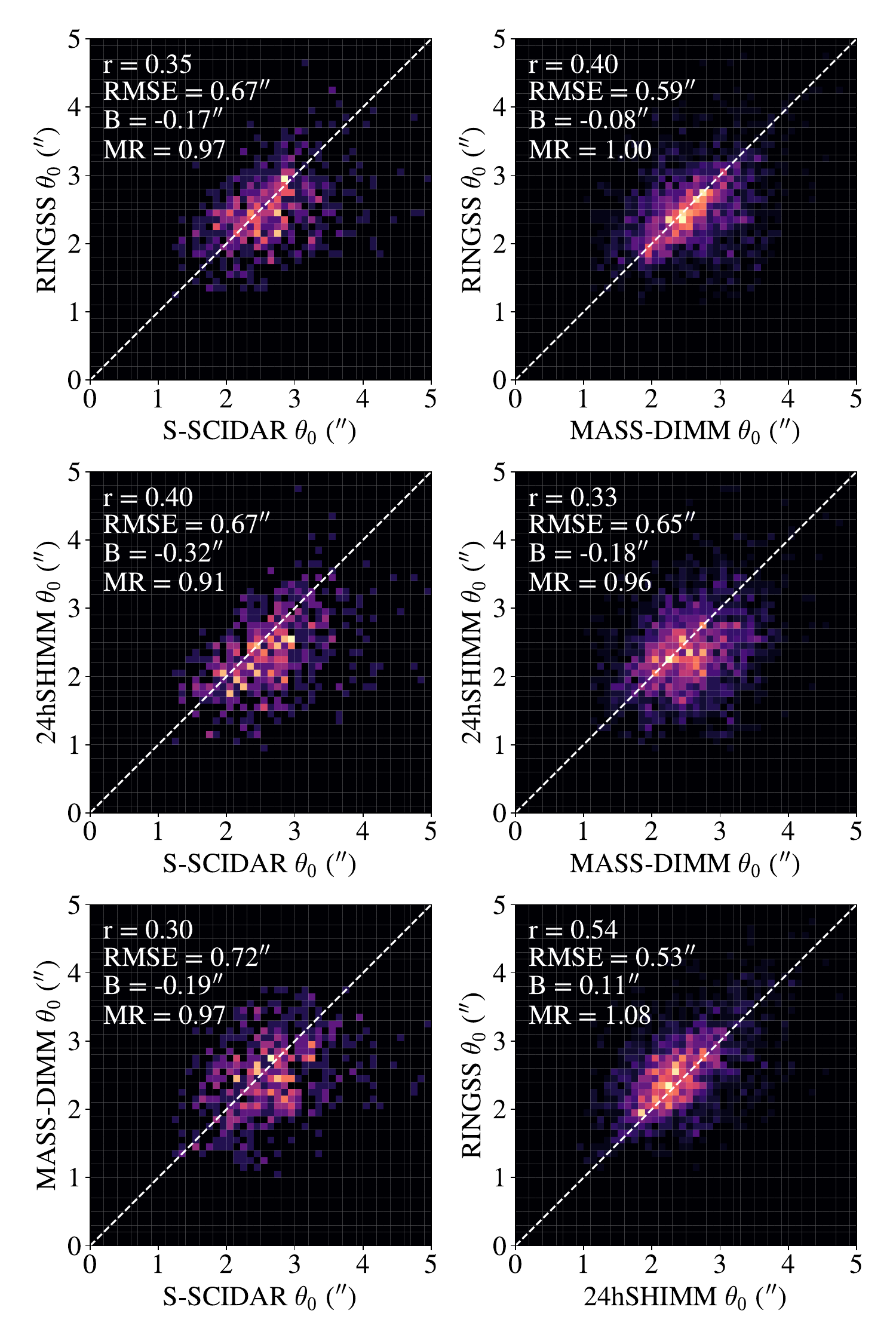}
\caption{Comparison of contemporaneous isoplanatic angle measurements during the campaign by the \gls{mass-dimm}, \gls{scidar}, \gls{shimm} and \gls{ringss}.}
\label{fig:iso}
\end{figure}

Figure~\ref{fig:iso} displays the comparisons of isoplanatic angle measured by all instruments. Unlike measurements of the seeing, it is observed that there is less correlation between all instruments. However, the variation of isoplanatic angle during the campaign was small. The strongest correlation, 0.54, is found between \gls{shimm} and \gls{ringss} which observed same targets, while other profilers sampled different turbulent volumes. The $h^{5/3}$ scaling in Eq. \ref{equation:isop} implies that this parameter is highly sensitive to  turbulence in the upper atmosphere. Therefore an accurate characterisation will require sensitivity to high-altitude turbulence. The \gls{shimm}, \gls{ringss} and \gls{mass} are limited in this regard by their response functions for the highest altitude layer which are several kilometres wide. The turbulence distributed over this layer will be averaged and reported at that height, leading to a reduction in accuracy. When taking optical propagation into account for observing at lower zenith angles, saturation of scintillation produced by the highest-altitude layers is an additional source of error for monitors based on weak-scintillation theory. The exception in this experiment being the \gls{ringss} and \gls{mass} which implement a correction process. This combination of factors is likely to explain the smaller correlation observed in measurements from the four instruments, while the median values agree fairly closely. 

%\begin{table}
% \caption{Summary of statistical comparison parameters for contemporaneous measurements of isoplanatic angle.}
% \label{tab:isoplanatic}
% \begin{tabular}{llcccc}
%  \hline
%   X - axis & Y - axis & r & RMSE & B & MR\\
%   & & & ($\arcsec$) & ($\arcsec$) &  \\
%   \hline
%  \gls{scidar} & \gls{ringss} & 0.35 & 0.69 & -0.17  & 0.97 \\
%  \gls{mass-dimm} & \gls{ringss} & 0.40 & 0.60 & -0.08 & 1.00 \\
%  \gls{scidar} & \gls{shimm} & 0.40 & 0.74 & -0.30 & 0.91 \\
%  \gls{mass-dimm} & \gls{shimm} & 0.33 & 0.68 & -0.17 & 0.97\\
%  \gls{scidar} & \gls{mass-dimm} & 0.30 & 0.75 & -0.19 & 0.97\\
%  \gls{shimm} & \gls{ringss} & 0.54 & 0.54 & 0.10 & 1.07\\
%  \hline
% \end{tabular}
%\end{table}

\subsection{Coherence time}\label{sec:coherencetime}

Knowledge of the coherence time is essential for \gls{ao} as it defines the minimum bandwidth of the system. The optical turbulence coherence time is typically on the scale of a few ms. It is related to the wind speed profile and turbulence strength in the following way \citep{Roddier1981},
\begin{equation} \label{eq:tau0}
    \tau_0 = 0.314 \frac{ r_0} {\overline{V}_{5/3}},
\end{equation}
where $\overline{V}_{5/3}$ is the weighted mean of the wind speed raised to the power of $5/3$,
\begin{equation} \label{eq:meanwind}
    \overline{V}_{5/3} = \left[\frac{\int^\infty_0 V(h)^{5/3} C_n^2(h) \, \mathrm{d}h}{\int^\infty_0 C_n^2(h) \, \mathrm{d}h}\right]^{3/5}.
\end{equation}

The instruments in this study employ a variety of strategies to measure the coherence time. The \gls{scidar} analyses the spatio-temporal cross-correlations of the scintillation measured in the pupil. Peaks that match atmospheric layers translate across the auto-covariance map with each successive time offset due to translation of the turbulent layers with wind.  The direction and speed of each of the layers is recorded and the mean wind speed calculated from  Eq.~\ref{eq:meanwind}. The \gls{scidar} is only able to directly estimate the wind speed of the strongest layers. Weak layers with no detected wind speed are assigned a value through interpolation of the measured wind speed profile. The \gls{shimm} takes a different approach, utilising the FADE method \citep{Tokovinin2008FADETime}, which involves fitting response functions, determined by layer wind speeds and \cn, to the measured temporal structure function of the Zernike defocus coefficient of the atmospheric wavefront distortions. The \gls{shimm} analysis differs slightly from the FADE instrument as wavefronts are reconstructed by the Shack-Hartmann yielding direct measurements of the Zernike defocus term, and only layer wind speeds need to be fitted. As the \gls{shimm} sampling rate was limited to 100Hz for this experiment, it was necessary to exclude 362 measurements that had a $\overline{V}_{5/3} > 15\;\mathrm{ms^{-1}}$ as the defocus structure function curve could not be sampled with a sufficient temporal resolution to fit a wind speed profile.
%The \gls{shimm} takes a different approach, utilising measurements of the $C_n^2 \mathrm{d} h$ profile, atmospheric wind speeds from the ECMWF ERA-5 reanalysis binned to the \gls{shimm} profile layers with the wind speed for the ground layer bin replaced by wind speed measured from a local weather station which typically has a higher temporal resolution. 
The \gls{mass-dimm} and \gls{ringss} utilise the method described in \citet{Kornilov2011StellarEvaluation} of including a wind shear component in the weighting functions, continuous exposures without gaps, and a fitting process to estimate the second moment of the wind $\overline{V}_2$ with the approximation of $\overline{V}_2 \approx 1.1 \overline{V}_{5/3}$ found by \cite{Kellerer2007AtmosphericMeasurement} enabling an estimate of the coherence time.

\begin{figure}
\centering
\includegraphics[width=\hsize]{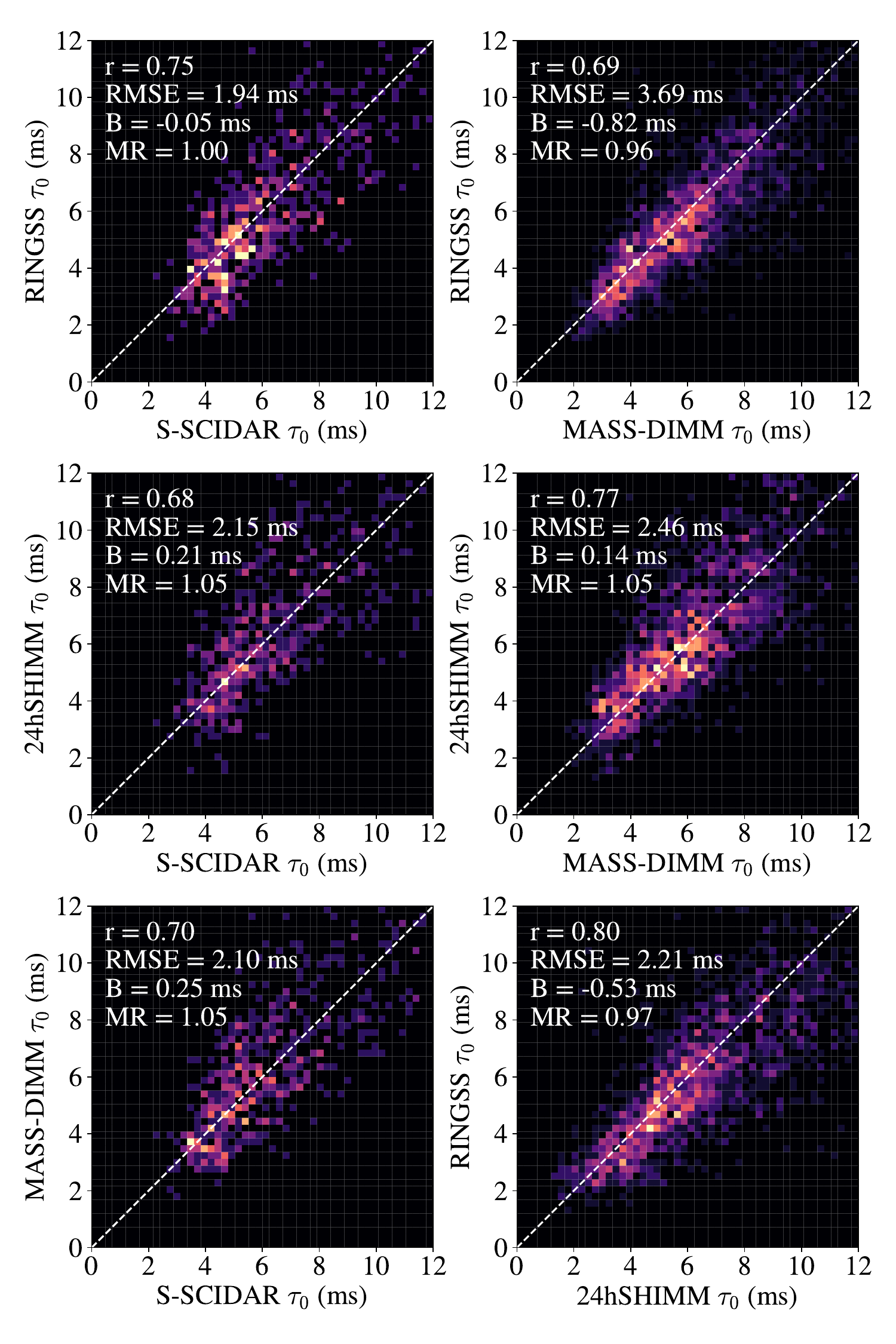}
\caption{Comparison of contemporaneous measurements of the atmospheric turbulence coherence time by the \gls{mass-dimm}, \gls{scidar}, \gls{shimm} and \gls{ringss}.}
\label{fig:coherencetime}
\end{figure}

Figure~\ref{fig:coherencetime} displays comparisons of coherence time measurements for the four instruments. The \gls{ringss} and \gls{mass} use the same method of calculating coherence time and agree strongly with little bias. The two instruments also agree well with the \gls{scidar}, again with little bias. The \gls{shimm} shows good correlation with all instruments too. The bias however is small but positive with respect to the \gls{scidar} and \gls{mass-dimm}. Lower elevation and imaging through more of the surface layer should lead to a negative bias, suggesting that the instrument may be overestimating coherence time which could be a result of the low frame rate. Finally, the lower correlation of some instruments with the \gls{scidar} may result from the fact that \gls{scidar} measures wind direction and corrects line-of-sight wind speed measurements to the wind speed parallel to the ground, which other instruments cannot do.
%{The \gls{shimm} technique however displays poorer agreement with the other three instruments which may be expected given the limited temporal resolution of forecast data. The coherence time is also biased towards smaller values }
%\textcolor{red}{with an overall median of 3.7~ms compared to the other instruments whose median coherence time measurements fall in the range 5.8-6.3~ms} which is 
%reflected in the median results given in Table~\ref{table:All}., however the correlation remains strong with the \gls{mass-dimm} and \gls{ringss}.

\subsection{Influence of wind direction}

\begin{figure}
    \centering
    \includegraphics[width=\hsize]{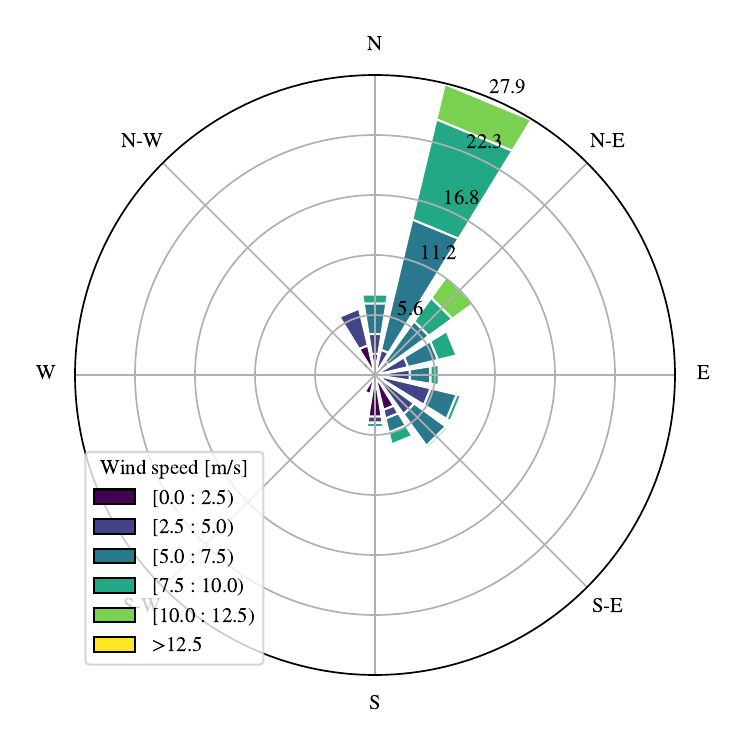}
    \caption{A wind rose displaying the distribution of wind speeds and directions measured 30~m above the ground by the Paranal meteo-tower for the six nights of the campaign.}
    \label{fig:windrose}
\end{figure}

Previous studies have observed that wake produced downwind of large telescope structures can have a significant effect on seeing conditions \citep{Sarazin1990VLT62}. Additionally, seeing at the 1998-\gls{dimm} tower has historically been stronger than that observed by the UTs for north-easterly and south-easterly winds \citep{Sarazin2008SeeingParanal}. A later study by \citet{Lombardi2010SurfaceObservatory} related this phenomenon to an increase in the strength of the surface layer. We therefore expect wind direction to influence the agreement between instruments in this campaign. The wind rose, figure~\ref{fig:windrose}, shows the distribution of wind speeds and directions measured 30~m above the ground by the meteo-tower between sunset and sunrise for all six nights of the campaign. The 30~m measurement is used over the 10~m measurement to minimise bias introduced by the Unit Telescopes (UTs) to the South and the VST to the SSW. The radial extent of the bars represents the fraction of the data with a given wind direction and it suggests, similar to previous studies such as \citet{Lombardi2009TheObservatory}, that it is mainly from the NNE.

\begin{figure}
    \centering
    \includegraphics[width=\hsize, trim = 0cm 0cm 0cm 0cm]{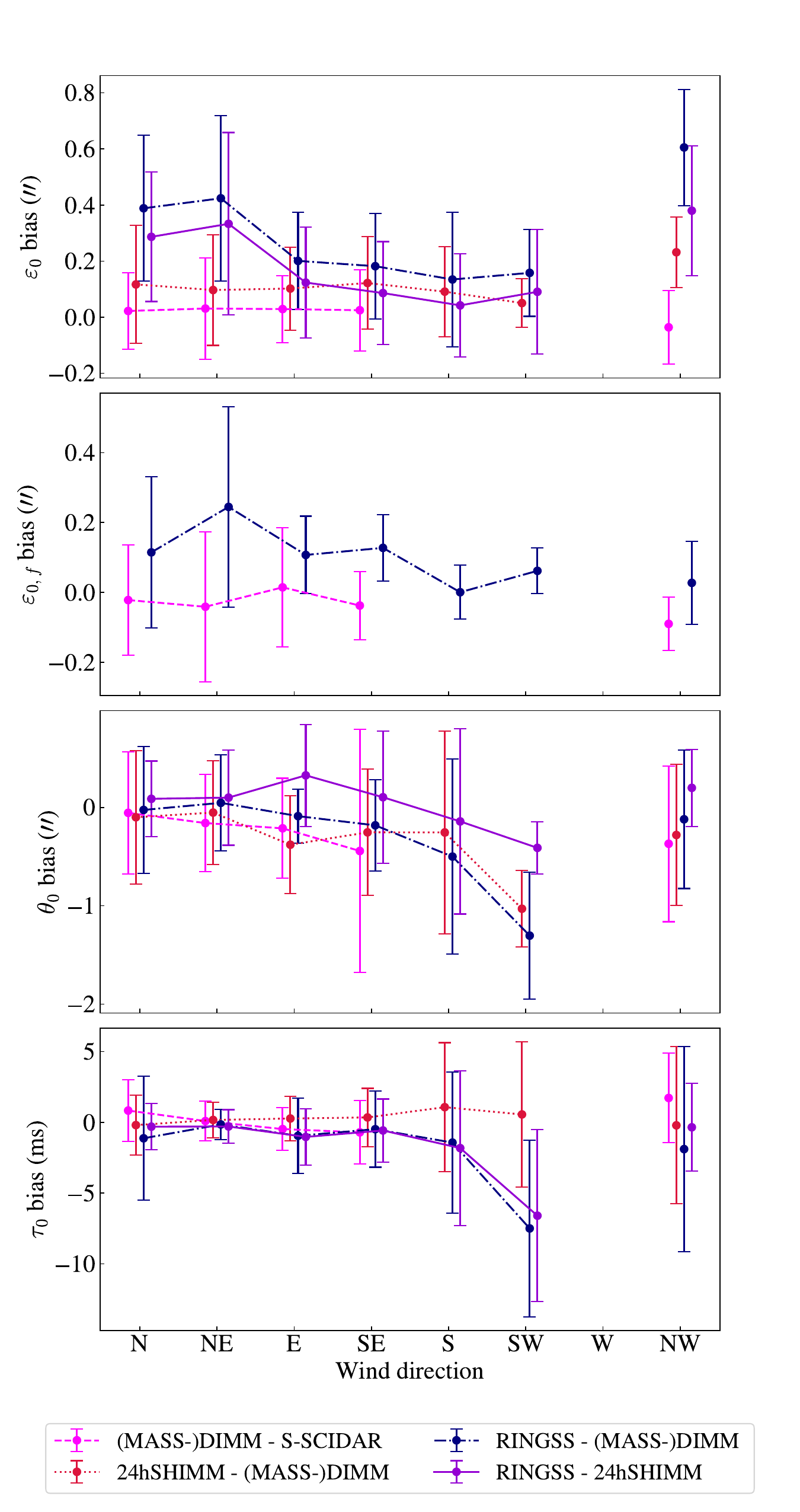}
    \caption{A plot showing the bias of measurements for all four integrated turbulence parameters, and the RMSE indicated by the error bars, as a function of wind direction for key pairs of instruments compared in this study. For the seeing, only \gls{dimm} data is used, but for other parameters the \gls{mass-dimm} results use the same line style. The legend indicates the Y - X instrument pair for which the bias and RMSE have been plotted.}
    \label{fig:wind_dirs}
\end{figure}

Figure~\ref{fig:wind_dirs} shows how the bias between pairs of instruments changes as a function of wind direction for eight directional bins. In addition, the error bars indicate the bias-corrected RMSE of the comparisons for each wind direction. Due to insufficient data for some wind directions, the correlation is not plotted. Additionally, there were no \gls{scidar} data points between South and West and insufficient data for all instruments for the West bin. These points have therefore been omitted. Seeing measurements during the campaign appear to be strongly influenced by wind direction. For instrument pairs other than the \gls{scidar} and \gls{mass-dimm}, the RMSE of instrument comparisons is larger for northerly winds. The \gls{ringss} bias appears sensitive to the wind direction with the largest bias corresponding to north-westerly winds, but the \gls{shimm} does not follow the same pattern - only seeing a larger bias compared to the \gls{mass-dimm} towards the North-West. However there are few data points for this bin. This figure does not take into account instrument pointing direction, which can also lead to discrepancy in measurements. As this sample of six nights is relatively small, the influence of pointing direction was investigated instead through analysing the median and standard deviation of seeing measured by the 2016-\gls{dimm} for all data in the ESO archive. This analysis showed a clear increase in median seeing for north-easterly and south-easterly winds for all pointing angles. Features strongly dependent on pointing angle included: larger variability at low elevation angles when the \gls{dimm} points SE and wind blows from the W and SW, and for the DIMM pointing SW while the wind blows from the North. The larger spread of data and bias for northerly winds experienced by the \gls{shimm} and \gls{ringss} may be related to their proximity to the edge of the platform, as shown in figure~\ref{fig:platform}, as air from the ground level will be driven up the mountain and mix with cooler air at the platform. By contrast, wind from the South will traverse the platform before reaching the \gls{shimm} and \gls{ringss}. The \gls{scidar} vs \gls{mass-dimm} seeing comparison has no identifiable dependence on wind direction which is expected as both instruments are raised above the ground and located away from the platform edges and buildings.

For the free atmosphere seeing and isoplanatic angle, dependence on wind direction at 30~m seems unlikely as both parameters are insensitive to ground layer turbulence. In reality, non-Kolmogorov turbulence in the surface layer which may arise from interaction of wind with buildings or heat sources can ``confuse'' turbulence monitoring instruments that expect a specific power spectrum (typically Von Karman or Kolmorogov), thus leading to inaccuracies in the characterisation of the turbulence profile that may depend on wind direction. Such effects are also encountered at low wind speeds and have been identified at the site by the \gls{slodar} \citep{Butterley2020CharacterizationSystem}. Figure~\ref{fig:windrose} shows that for southerly winds, a wind speed of less than 3~ms$^{-1}$ is proportionally more frequent. For the coherence time, which is also dependent on the vertical wind speed profile, the biases are small relative to the spread of the data, except for the SW which may result from a small number of samples. The wind direction does not seem to have a significant influence on the bias or RMSE of these comparisons, however there is a trend towards a larger negative bias for most instrument comparisons in the NE to SW section of the graph. A full treatment of wind directional discrepancies at Paranal would require a significantly larger data set and is beyond the scope of this study.
%Fig ure~\ref{fig:wind_dirs} indicates that there is no statistically significant change in the bias for the isoplanatic angle and free atmosphere seeing, which is expected as the wind direction is likely to have the greatest influence on local turbulence. The greatest bias in the seeing appears to coincide with northerly winds - for which the \gls{ringss} and \gls{shimm} record consistently stronger seeing than the other two instruments, although the statistical significance is low due to inherently large spread in the data and a smaller number of data points in some directions as indicated by figure~\ref{fig:windrose}. Referring to the site map, figure~\ref{fig:platform}, a Southerly wind would blow across the platform and past the VST before reaching the \gls{shimm} and \gls{ringss}. On the other hand a northerly wind blows up the side of the mountain, and these two instruments were located very close to the edge of the platform. Further study with well-calibrated instruments would be needed to investigate the root cause of this effect, however we speculate that the increased mixing of warmer air driven up the mountainside with cooler air at the platform may be leading to increased surface layer strength near to the edge of the platform. For the coherence time we are unable to make similar conclusions as this requires disentangling the contribution of the wind velocity profile from integrated turbulence profile; this is beyond the capability of instruments other than the \gls{scidar}.}

%\begin{table}
% \caption{Summary of statistical comparison parameters for contemporaneous measurements of coherence time.}
% \label{tab:coherence_time}
% \begin{tabular}{llcccc}
%  \hline
%   X - axis & Y - axis & r & RMSE & B & MR\\
%   & & & (ms) & (ms) &  \\
%    \hline
%  \gls{scidar} & \gls{ringss} & 0.75 & 1.94 & -0.05  & 1.00 \\
%  \gls{mass-dimm} & \gls{ringss} & 0.69 & 3.78 & -0.82 & 0.96 \\
%  \gls{scidar} & \gls{shimm} & 0.68 & 3.16 & -2.60 & 0.62 \\
%  \gls{dimm} & \gls{shimm} & 0.73 & 4.04 & -3.04 & 0.62\\
%  \gls{scidar} & \gls{dimm} & 0.70 & 2.11 & 0.25 & 1.05 \\
%  \gls{shimm} & \gls{ringss} & 0.82 & 3.30 & 2.48 & 1.62 \\
%  \hline
% \end{tabular}
%\end{table}

\subsection{Optical turbulence profiles}\label{sec:otp}

Optical turbulence profiles are characterised by the refractive index structure constant $C_n^2$ as a function of vertical height above the ground. The instruments in this study record the sum of $C_n^2$ over a given volume $\mathrm{d}h$ for each layer using an inversion process. To facilitate a comparison between all instruments which use different models and layers, the \gls{ringss}, \gls{mass-dimm} and \gls{shimm} are directly compared with the high-resolution \gls{scidar} profiles through binning using instrument response functions.

The response functions dictate the measured \cn response to a single, thin turbulent layer placed at any height throughout the atmosphere. These functions are typically evaluated in simulation by passing a single, thin layer from the ground to the upper atmosphere and plotting the \cn measured by the instrument in each altitude bin. For scintillation-based instruments such as \gls{ringss}, \gls{scidar} and \gls{mass} the response functions usually manifest as triangles on a log scale of height, centred on the altitude of the turbulent layer reconstructed and crossing adjacent bins at half of the input turbulence strength \citep{Tokovinin2003RestorationIndices,Tokovinin2021MeasurementImages}. 

\begin{figure}
\centering
\includegraphics[width=\hsize, trim= 2cm 13.5cm 2cm 5cm]{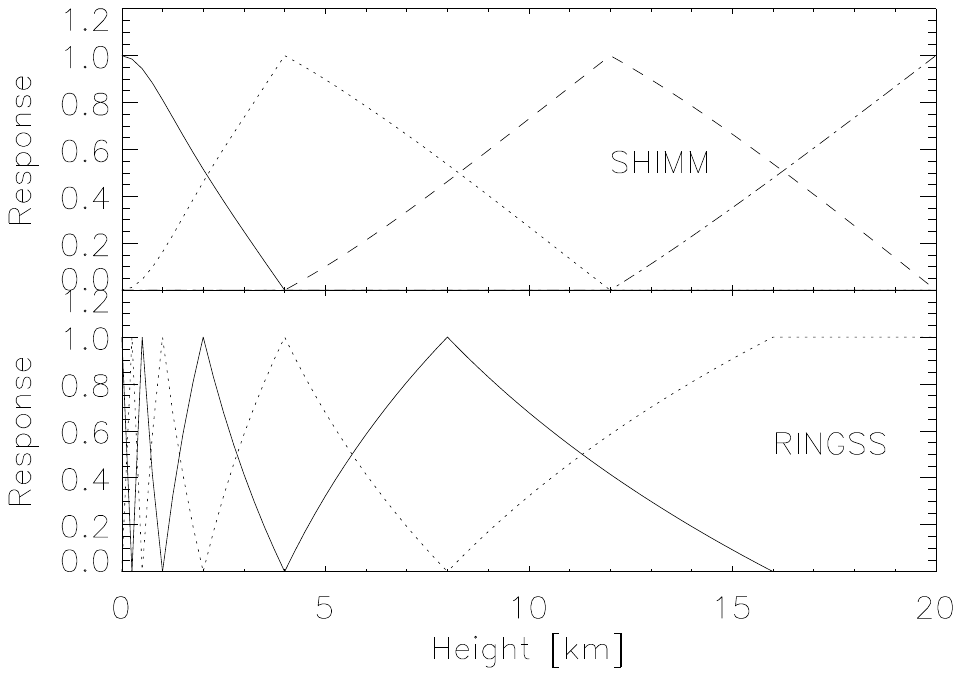}
\caption{A plot of the response functions for the \gls{shimm} and \gls{ringss}. The alternating line styles differentiate the response functions of each reconstructed layer. The sum of responses from all layers is approximately one.}
\label{fig:response}
\end{figure}

For the \gls{shimm}, this approximation also holds well, except for between the ground layer and the first layer. The response functions $f_i(h)$ for the \gls{shimm} and \gls{ringss} are displayed in figure~\ref{fig:response} on a linear scale of height. These instruments, as well as  \gls{mass}, estimate turbulence strength in discrete layers as $C_n^2(h_i){\rm d}h = \int f_i(h) C_n^2(h) \rm{d}h$. The response functions for the \gls{mass} can be found in \cite{Kornilov2003MASS:Distribution}.

\begin{figure}
\centering
\includegraphics[width=\hsize]{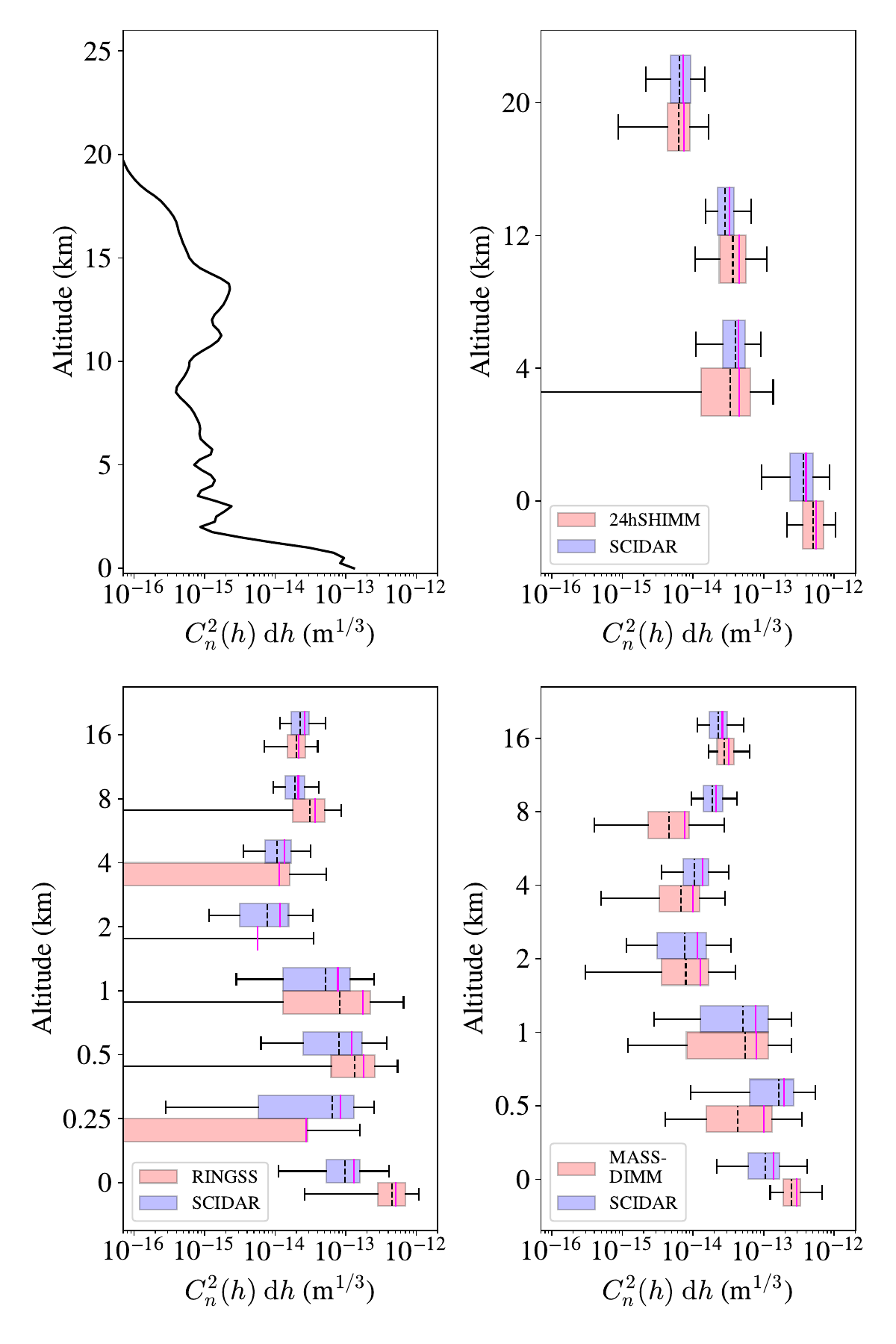}
\caption{A comparison of \cn profile measurements for all instruments with contemporaneous measurements from the \gls{scidar}. The red boxes show the instrument data from each fitted layer, and the adjacent blue boxes the contemporaneous measurements (within +/- two minutes) from the \gls{scidar} which have been binned to match the instrument layers using the response functions. The extent of coloured boxes represents the first and third quartiles, the dashed line the median measurement, the magenta line the mean, and the whiskers the fifth and 95th percentiles of the distribution. From top left to bottom right, the plot shows the mean \gls{scidar} profile, and box and whisker plots for the \gls{shimm}, \gls{ringss} and \gls{mass-dimm} compared with \gls{scidar}. Significantly smaller values in the top-left panel, compared to other panels, are explained by the thinner ${\rm d}h=0.25$\,km layers of the \gls{scidar} profiles.}
\label{fig:profiiles}
\end{figure}

Figure~\ref{fig:profiiles} displays a box and whisker plot of optical turbulence profile measurements from the \gls{shimm}, \gls{ringss} and \gls{mass-dimm} compared with contemporaneous \gls{scidar} profiles. The \gls{scidar} profiles have been binned down to the instrument layers using the response functions and only data within $\pm 2$ minutes of an \gls{scidar} measurement have been used. The whiskers represent the 5th and 95th percentiles of the distribution, the median is shown as a dashed black line and the mean as a solid magenta line. It is therefore possible to simultaneously compare mean profiles and distributions of measurements in individual layers. Figure~\ref{fig:profiiles} indicates that all instruments measure a significantly stronger ground layer than the equivalent \gls{scidar} measurement.

% REMOVED It is notable that there is a strong discrepancy in the \gls{mass-dimm} ground layer where we might expect a better agreement as both instruments are at the same height. However, this could be explained partially by the dome turbulence subtraction implemented by the \gls{scidar} and also by underestimation in the 0.5~km \gls{mass-dimm} layer.

A notable feature of the \gls{mass-dimm} profile is a significant underestimation in the 8~km layer, which appears to be the driving cause of the smaller value of median free-atmosphere seeing. For \gls{ringss} and the \gls{shimm}, some layers register zero \cn, hence anomalous boxes and whiskers such as the 4km layer for the \gls{shimm} and 2~km layer in \gls{ringss} on a a log-scale of \cn. Mean values however agree well for the free-atmosphere layers.

%\begin{figure}
%\centering
%\includegraphics[width=\hsize]{figures/mean_profiles.pdf}
%\caption{A comparison of mean profiles from measurements contemporaneous with the \gls{scidar}. The red points represent the mean instrument profile and the black points the mean of \gls{scidar} profiles binned by the corresponding instrument response function. The top left plot shows the mean 100-layer \gls{scidar} profile, the top right the \gls{shimm} result, bottom left \gls{ringss} and bottom right \gls{mass-dimm}.}
%\label{fig:profiiles}
%\end{figure}

\begin{figure}
\centering
\includegraphics[trim=2.4cm 13cm 2.3cm 5cm, width=\hsize]{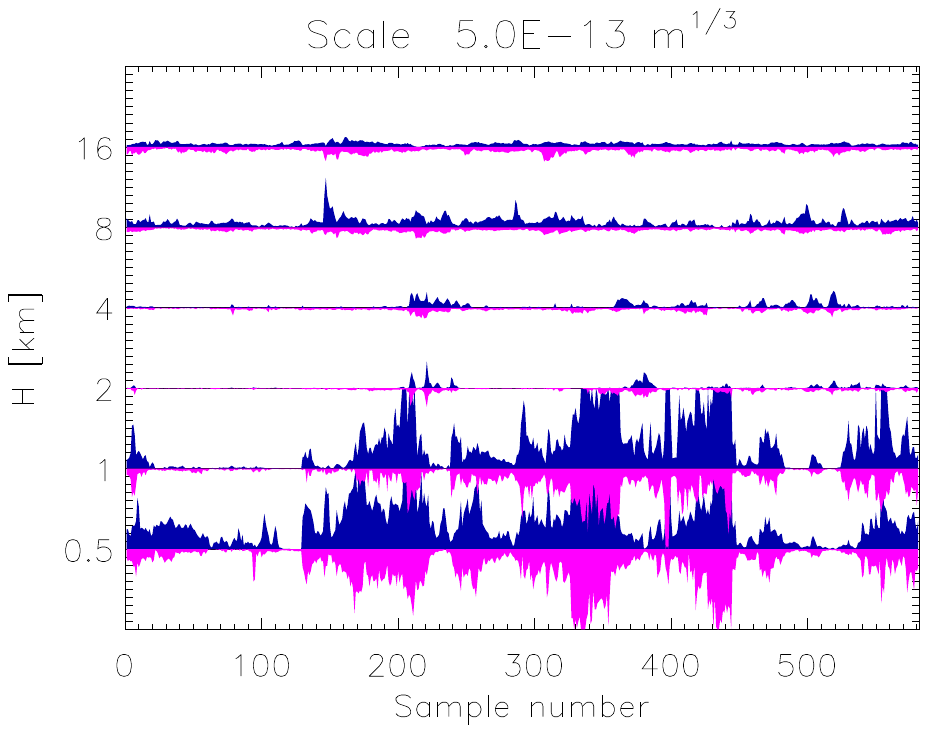}
\caption{Turbulence profiles measured simultaneously by \gls{ringss}
  (up-facing blue bars) and
  \gls{scidar} (down-facing magenta bars). \gls{scidar} is matched in resolution
  and time   to \gls{ringss} with the sample number indicating the nth \gls{scidar} measurement taken during the campaign. The width of each band is $5\times10^{-13} \mathrm{m}^{1/3}$.}
\label{fig:ringss-scidar}
\end{figure}

\begin{figure}
    \centering
    \includegraphics[trim=2.4cm 13cm 2.3cm 5cm,width=\hsize]{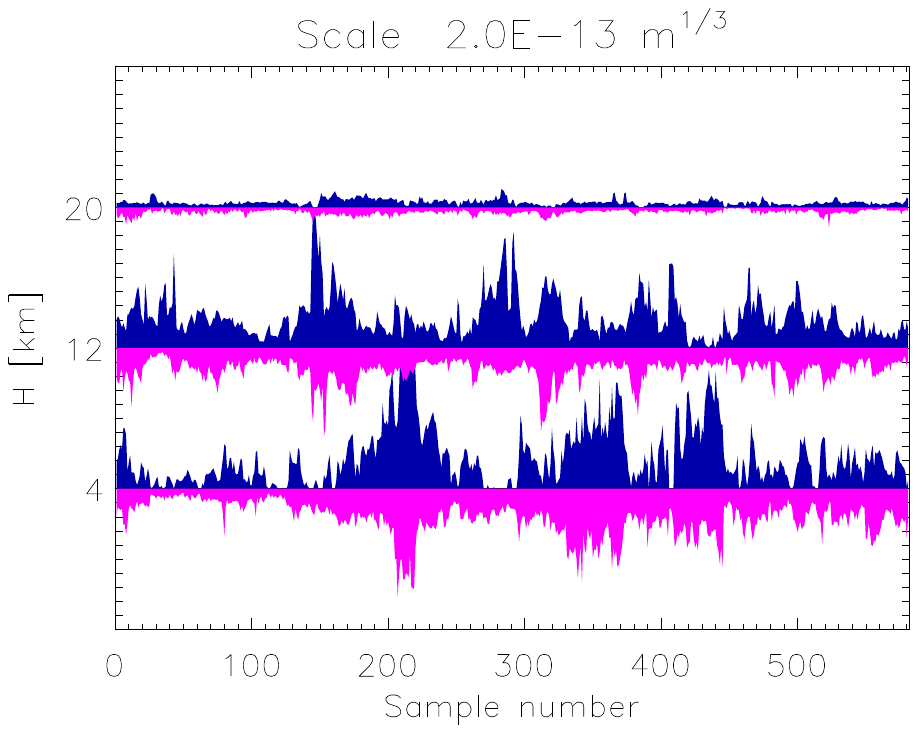}
    \caption{Turbulence profiles measured simultaneously by \gls{shimm}
  (up-facing blue bars) and
  \gls{scidar} (down-facing magenta bars). \gls{scidar} is matched in resolution
  and time to \gls{shimm} with the sample number indicating the nth \gls{scidar} measurement taken during the campaign. The width of each band is $2\times10^{-13} \mathrm{m}^{1/3}$.}
    \label{fig:shimm-scidar}
\end{figure}

Figure~\ref{fig:ringss-scidar} shows a detailed comparison between vertical turbulence profiles measured by \gls{ringss} with all 611 available \gls{scidar} profiles matched in time and resolution. Despite different locations and different target sources, we note a strong agreement of timing and localisation of strong turbulence packets, especially in the 0.5 and 1-km layers. The ground layer is not included in this comparison. Figure~\ref{fig:shimm-scidar} shows a similar plot for the \gls{shimm}. It suggests that the correlation between lower-altitude layers is higher than for high-altitude layers, evidencing the low correlation in isoplanatic angle.

\subsection{Day and night measurements}\label{sec:24hr}

The \gls{shimm} measures \gls{ot} profiles continuously for 24-hours a day by operating at short-wave infrared wavelengths. Compared to the visible light, this extends the validity of the weak-scintillation assumption and reduces the sky background. Additional techniques for rapid background subtraction \citep{Griffiths2023} are also employed to ensure accurate photometry. 

\begin{figure}
\centering
\includegraphics[width=\hsize]{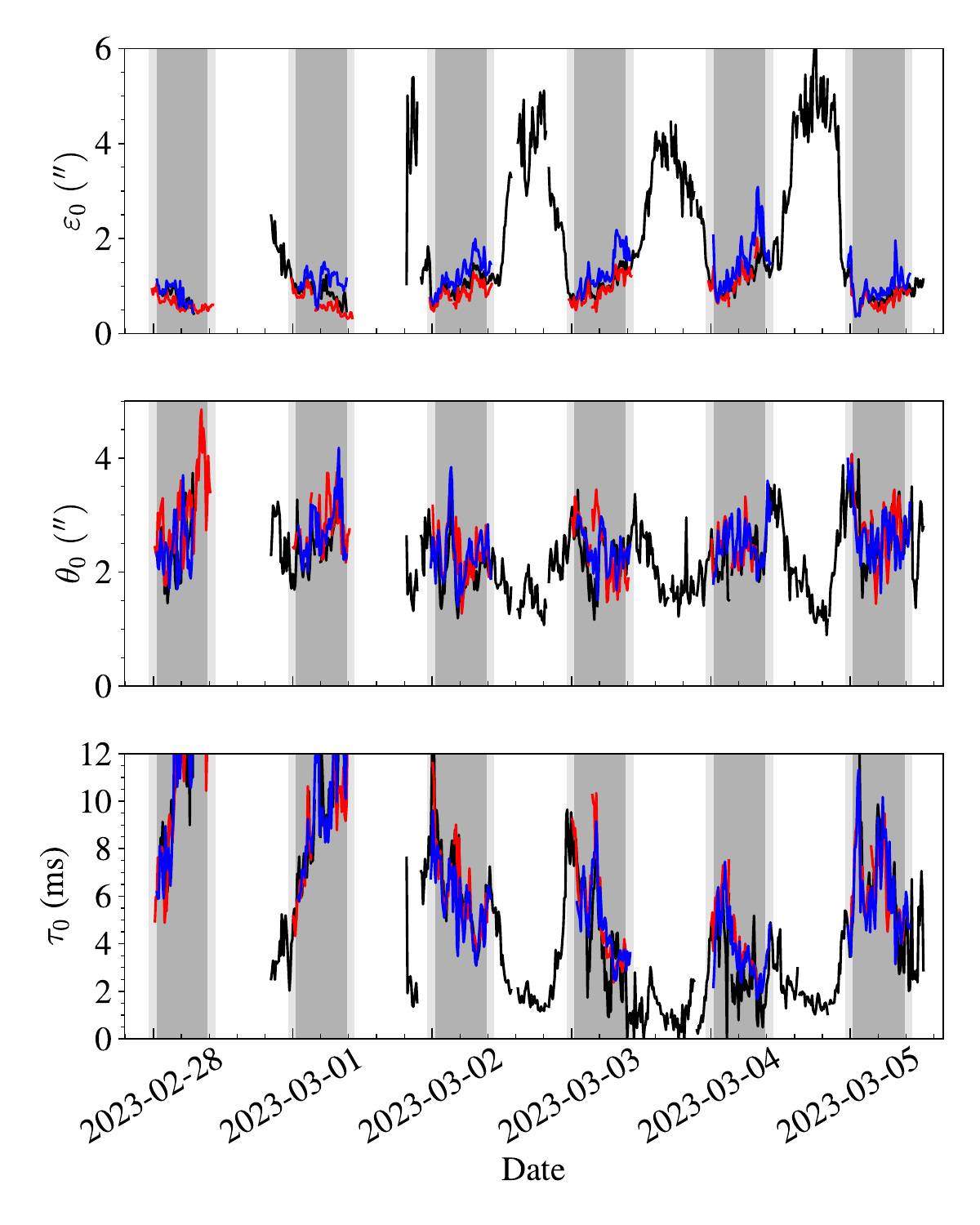}
\caption{Integrated parameters measured by the \gls{shimm} during the campaign. The black line represents \gls{shimm} measurements, the red line \gls{dimm} measurements for seeing and \gls{mass-dimm} for the coherence time and isoplanatic angle, and the blue line the \gls{ringss}. All data sets have been binned to 10-minute intervals for presentation and dates are in UTC. The white, grey and light grey shades of the background represent daytime, night and twilight respectively}
\label{fig:24hr}
\end{figure}

Figure~\ref{fig:24hr} shows a continuous plot of the three main integrated turbulence parameters estimated by the \gls{shimm}: seeing, isoplanatic angle and coherence time. Because the instrument produced a measurement every 1-2 minutes, for presentation purposes the data have been binned such that each data point represents the average of any measurements that fall into ten-minute bins. The sharp diurnal variation in seeing is immediately evident from the graph, with a repetitive, sharp drop in the seeing after sunset leading to the best conditions in earliest part of the night. The general trend thereafter appears to be a gradual increase in the seeing until just after sunrise where it rises very strongly. More work is needed to understand the underlying processes behind this behaviour and the influence of meteorological parameters.

The median value of the daytime seeing, calculated between sunrise and sunset, was found to be 2.65$\arcsec$, isoplanatic angle 2.05$\arcsec$ and coherence time $2.4$~ms. It is notable that measurements of the isoplanatic angle, which is insensitive to low-altitude turbulence, do not experience the same distinct variation. This suggests that the increased turbulence strength during daytime is a result of solar heating at the ground affecting the boundary layer, and the upper atmosphere is relatively unaffected. The coherence time follows a similar trend to the Fried parameter likely due to dominance of the strong ground layer turbulence.

%-------------------------------------- Two column figure (place early!)
%   \begin{figure*}
%   \centering
   %%%\includegraphics{empty.eps}
   %%%\includegraphics{empty.eps}
   %%%\includegraphics{empty.eps}
%   \caption{Adiabatic exponent $\Gamma_1$.
%               $\Gamma_1$ is plotted as a function of
%               $\lg$ internal energy $\mathrm{[erg\,g^{-1}]}$ and $\lg$
%               density $\mathrm{[g\,cm^{-3}]}$.}
%              \label{FigGam}%
%    \end{figure*}
%

%
%                                                One column figure
%----------------------------------------------------------------- 
%   \begin{figure}
%   \centering
%   %%%\includegraphics[width=3cm]{empty.eps}
%      \caption{Vibrational stability equation of state
%               $S_{\mathrm{vib}}(\lg e, \lg \rho)$.
%               $>0$ means vibrational stability.
%              }
%         \label{FigVibStab}
%   \end{figure}
%-----------------------------------------------------------------
\section{Conclusions}

An optical turbulence monitoring campaign has been carried out at Cerro Paranal observatory between the 27th February and 5th March 2023. The aim of this study was to characterise novel turbulence monitoring instruments, the \gls{shimm} and \gls{ringss}, against existing instruments at the site through comparison measurements of vertical \gls{ot} profiles and integrated parameters including the seeing, free-atmosphere seeing, isoplanatic angle and coherence time.

Data collected from these two instruments during the campaign were further compared against measurements from the \gls{scidar} and the \gls{mass-dimm} by assessing the RMSE, bias and correlation of contemporaneous data from pairs of instrument. Additionally median values from the whole campaign were calculated and compared to long-term averages.

It was found, as in previous campaigns, that the seeing measured near the old 1998-DIMM tower was significantly larger than for the \gls{scidar} and 2016-\gls{mass-dimm}. In general, however, strong correlation was found across all seeing and free-atmosphere seeing measurements. Isoplanatic angle measurements displayed a close agreement in median values, but were less correlated between all instruments, which is likely a result of limitations in sensitivity to high altitude turbulence and differences in the sampled turbulence volumes. Coherence time measurements were strongly correlated between all instruments, however the RMSE of distributions was relatively large.
%although the \gls{shimm} technique of using wind speed measurements from meteorological forecast data appears to strongly underestimate the coherence time. 
The influence of wind direction on statistical agreement between measurements was also investigated which showed increased spread and bias in \gls{ringss} and \gls{shimm} seeing comparisons with the \gls{mass-dimm} for northerly winds. Additionally, changes in bias for parameters that should have no dependence on the wind direction could be attributed to non-Kolmogorov effects.

The accuracy of \gls{ot} profiling was also investigated by comparison of profiles with contemporaneous \gls{scidar} measurements binned using instrument response functions. The two visiting instruments were found to agree well with the \gls{scidar}, with expected bias towards stronger turbulence in the ground layer. It was also observed that the \gls{mass-dimm} systematically underestimates the 8~km layer.

Finally, the first measurements of continuous optical turbulence parameters at Paranal were presented which indicate a predictable and extreme diurnal variation in seeing with a median daytime value of 2.65$\arcsec$ compared to equivalent night-time median of 0.88$\arcsec$, which is assumed to be driven by changes in the boundary layer due to solar heating in the early morning and rapid cooling in the evening as similar changes are not present in the isoplanatic angle which is sensitive to high altitude turbulence. This experiment suggests that the best seeing conditions are in the earliest part of the night.

\section*{Acknowledgements}

The turbulence monitor field campaign at Paranal was performed under the resources allocated by ESO in the Technical Time Request TTR-110.0010. The authors would also like to thank Jose Velasquez for operating the Stereo-SCIDAR during the experiment, and  AO and ML for their hard work in organising and hosting the campaign. RG acknowledges his Science and Technology Facilities Council studentship 2419794 and funding for the \gls{shimm} development under UK Research and Innovation (MR/S035338/1). The authors would like to thank Dr. Tony Travouillon for his review which improved the clarity and completeness of the analysis.

%%%%%%%%%%%%%%%%%%%%%%%%%%%%%%%%%%%%%%%%%%%%%%%%%%
\section*{Data Availability}

\gls{ringss}, \gls{shimm} and \gls{scidar} data may be found in the accompanying supplementary files. \gls{mass-dimm} data is available from the \gls{eso} \gls{asm} archive \footnote{http://archive.eso.org/cms/eso-data/ambient-conditions.html}.
% Data availability RINGSS
 
%The inclusion of a Data Availability Statement is a requirement for articles published in MNRAS. Data Availability Statements provide a standardised format for readers to understand the availability of data underlying the research results described in the article. The statement may refer to original data generated in the course of the study or to third-party data analysed in the article. The statement should describe and provide means of access, where possible, by linking to the data or providing the required accession numbers for the relevant databases or DOIs.



\bibliographystyle{mnras}
\bibliography{references_old} % if your bibtex file is called example.bib

%\begin{thebibliography}{99}
%\end{thebibliography}

%%%%%%%%%%%%%%%%%%%%%%%%%%%%%%%%%%%%%%%%%%%%%%%%%%

%%%%%%%%%%%%%%%%% APPENDICES %%%%%%%%%%%%%%%%%%%%%

\appendix

\section{Statistical comparison parameters}\label{sec:appendA}

In this section, the equations for statistical comparison parameters used in figures 2-5 and tables 1-2 are defined. In all equations $i = {1,2,3 \dots N}$ indicates a sample of $N$ independent turbulence parameter measurements, $X_i$ the measurement of the parameter by instrument $X$ and $Y_i$ the contemporaneous measurement of the parameter by instrument $Y$. The bias, $\mathrm{B}$ is defined as

\begin{equation}
\mathrm{B} = \sum_{i=1}^N \frac{ Y_i - X_i} {N},   
\end{equation}

the root mean square error (with bias subtracted), or $\mathrm{RMSE}$, as

\begin{equation}
\mathrm{RMSE} = \sqrt{ \sum_{i=1}^N \frac{ ([Y_i -\overline{Y_i}] - [X_i - \overline{X_i}])^2} {N} },
\end{equation}

where $\overline{Y_i}, \overline{X_i}$ are the means of the contemporaneous measurements, and the mean ratio by

\begin{equation}
    \mathrm{MR} = \frac{1}{N}\sum_{i=1}^N \frac{Y_i}{X_i}.
\end{equation}
%%%%%%%%%%%%%%%%%%%%%%%%%%%%%%%%%%%%%%%%%%%%%%%%%%

% Don't change these lines
\bsp	% typesetting comment
\label{lastpage}
\end{document}